\documentclass[12pt, prx,  aps, unsortedaddress,superscriptaddress,floatfix,amsmath,amssymb]{revtex4-1}
\usepackage{bm}
\usepackage{amsmath, amsfonts, amssymb}
\usepackage{graphics, mathrsfs,bm}
\usepackage{pstricks}
\usepackage{graphicx}
\usepackage[pdftex, bookmarks={false}, pdfauthor={Rudro R Biswas}, pdftitle={TGR}]{hyperref}
\usepackage{subfigure}
\allowdisplaybreaks[1]

\newcommand\ba{\begin{array}}
\newcommand\ea{\end{array}}
\newcommand\nn{\nonumber}
\newcommand\ri{\right}
\renewcommand\le{\left}

\newcommand{\feyn}[1]{#1\kern-0.45em/}

\renewcommand\d{\delta}

\newcommand\f{\phi}

\renewcommand\k{\kappa}

\newcommand\n{\nu}

\newcommand\p{\pi}

\newcommand\rr{\rho}

\newcommand\mbx{\mbs{x}}

\newcommand\y{\eta}

\newcommand\mc{\mathcal}

\newcommand\mbs{\boldsymbol}

\begin{document}
\title{Gravitational Response of Topological Quantum States of Matter}
\author{Guodong Jiang}
\affiliation{Department of Physics and Astronomy, Purdue University, West Lafayette, IN 47907.}
\author{YingKang Chen}
\affiliation{Department of Physics and Astronomy, Purdue University, West Lafayette, IN 47907.}
\author{\textcolor{black}{Srividya Iyer-Biswas}}
\affiliation{Department of Physics and Astronomy, Purdue University, West Lafayette, IN 47907.}
\affiliation{Santa Fe Institute, Santa Fe, NM 87501.}
\author{Rudro R. Biswas}
\email{rrbiswas@purdue.edu}
\affiliation{Department of Physics and Astronomy, Purdue University, West Lafayette, IN 47907.}

\begin{abstract}
Identifying novel topological properties of topological quantum states of matter, such as exemplified by the quantized Hall conductance, is a valuable step towards realizing materials with attractive topological attributes that guarantee their imperviousness to realistic imperfections, disorder and environmental disturbances. Is the gravitational coupling coefficient of topological quantum states of matter a promising candidate? Substantially building on well established results for quantum Hall states, \textcolor{black}{using disclinations as tools for controlled creation of pristine spatial curvature  free of undesirable artifacts such as would interfere with the electronic motion of interest,} herein we report that a large class of lattice topological states of matter exhibit gravitational response, i.e., charge response to \textcolor{black}{\emph{intrinsic}} spatial curvature. \textcolor{black}{This phenomenon} is characterized by a \emph{topologically} quantized coupling constant. Remarkably, the charge-gravity relationship remains \emph{linear} in the curvature, up to the maximum curvature achievable on the lattice, demonstrating absence of higher order nonlinear response. Our findings facilitate articulating the physical principles underlying the topological quantization of the gravitational coupling constant, in analogy with the insights offered by the Chern number description of the quantized Hall conductance. 
\end{abstract}

\maketitle

Fundamental aspects of the interplay between spacetime and geometry have been of longstanding interest in multiple contexts in physics -- for instance, Einstein's equation in general relativity encodes once such relation, while the interplay between nematics and real space curvature is a rich and frontier area of soft matter research\cite{2022-zhang-mr,2014-keber-bq,2009-bowick-id,2007-santangelo-cg,2006-vitelli-zx,2002-nelson-ci}. Exotic topological quantum states of matter, of which quantum hall states are paradigmatic, offer a rich playground for exploring universal responses of quantum geometry to external perturbations. The central question of interest here is:  Are there universal aspects to how quantum states of matter respond to real space curvature, i.e., gravitational perturbations? Moreover, are there transcendent insights equally applicable to gravitational responses of topological quantum and classical states of matter?

\begin{figure}[t]
\begin{center}
\resizebox{0.8\textwidth}{!}{\includegraphics{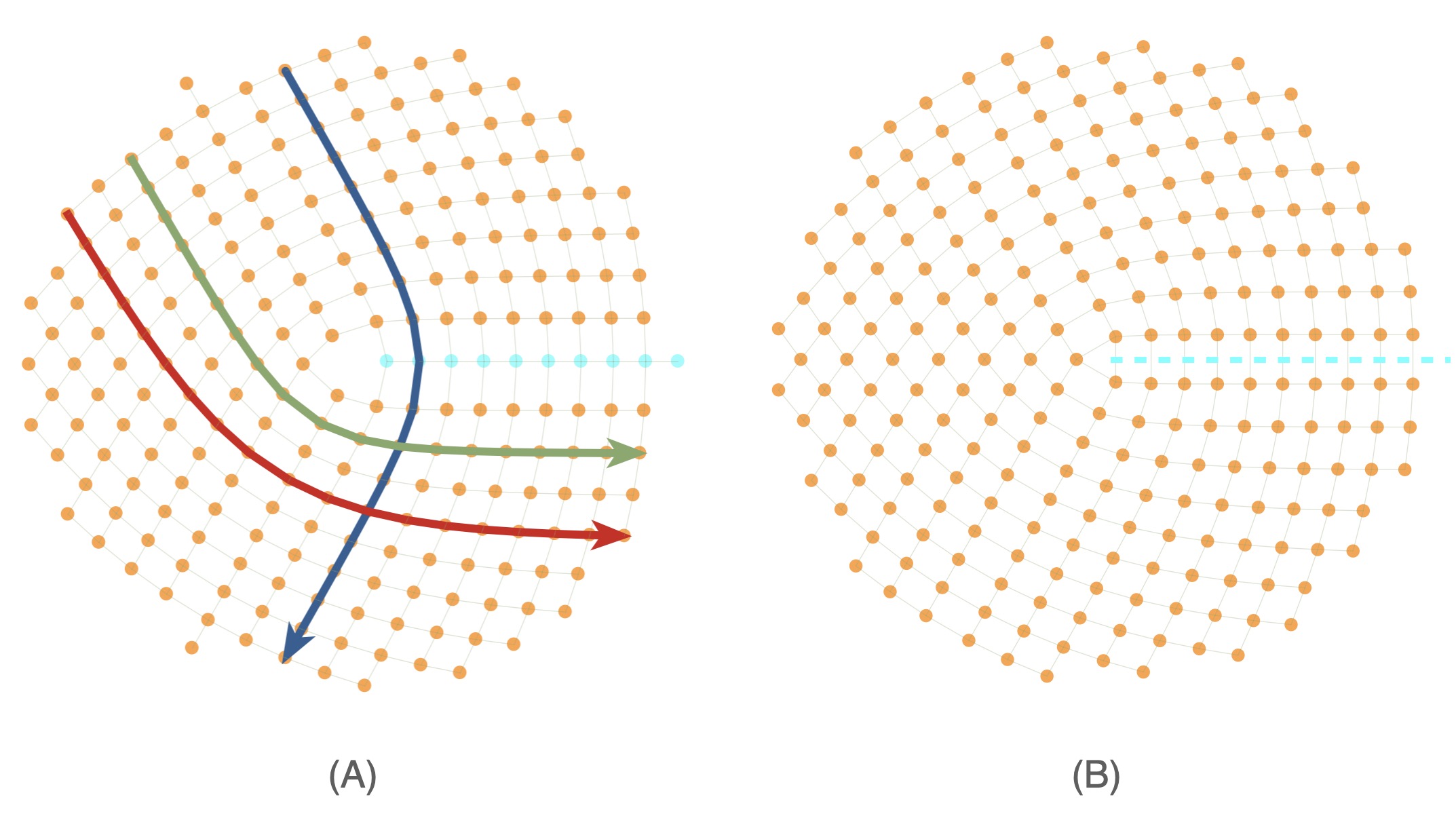}}
\caption{\emph{Disclinations as instantiations of spatial curvature.} \textcolor{black}{Distinct instantiations of three-fold disclinations on a square lattice with the \emph{same} intrinsic Gaussian curvature, centered at: (A) a plaquette center, and (B) a lattice site. The disclination induces geometric curvature: All three curved arrows in (A), which are geodesics, i.e., `straight lines' locally at all natural coordinate systems along their paths, start out parallel; however, geodesics traversing the disclination on opposite sides stop being parallel, in violation of Euclid's fifth postulate and a hallmark of curved space, intersecting further on with an angle of $\p/2$. This angle is the (dimensionless) integrated curvature of the disclination, also its Frank angle, and is quantized due to the rotational symmetry of the lattice. The two disclinations differ by a row of atoms (dislocation), indicated by the blue sites in (A) and dashed line in (B). Topological electronic states at such distinct classes of disclinations display differing electronic phenomena in general\cite{2013-gopalakrishnan-lk}, yet we find that they obey the same topological gravitational physics. We also find such universality in the presence of varying extrinsic curvatures, generated by differing strain fields (Figure~\ref{fig-strained}).}}
\label{fig-disclinations}
\end{center}
\end{figure}

Some valuable clues are provided by the well known Wen and Zee results for continuum quantum Hall states subjected to small and gently varying spatial curvature\cite{1992-wen-fk}, wherein the linear response coefficient characterizing local charge density response to Gaussian curvature is quantized. More recently, the result has been generalized to quantum Hall states on a lattice subjected to singular curvatures\cite{2016-biswas-kx}. However, unlike the celebrated example of the quantized Hall conductance, the quest to establish a Chern-number like conceptual framework for this (gravitational) coefficient in terms of topological properties of the electronic wavefunction has proved elusive. Here, we advance the goal of elucidating the quantum topological basis for quantization of charge response to spatial geometry by establishing the phenomenology, existence  and unifying relations characterizing quantized curvature-charge response in a large class of two dimensional lattice topological phases.

\textcolor{black}{We consider gapped two dimensional topological phases with rotational symmetry (but which do not require rotational symmetry to be defined), e.g., a Chern insulator on a lattice with rotational symmetry. Their gravitational response can be exposed by their behavior at disclinations since, in the presence of appropriate rotational symmetry, disclinations act as locations with well-defined integrated intrinsic Gaussian curvature equal to the Frank angle of the disclination\cite{2016-biswas-kx} (Figure~\ref{fig-disclinations}). Disclinations with the same Frank angle, i.e., the same net intrinsic Gaussian curvature, can have significant differences, such as possessing different extrinsic Gaussian curvatures depending how they are allowed to relax\cite{2022-vafa-sf}, or belonging to differing classes\cite{2013-gopalakrishnan-lk} based on, for e.g., dislocation parity. Yet, we demonstrate in this paper that given a gapped topological phase on a lattice,  constructed from Chern insulators, the charge accumulated at any disclination is a function solely of the net intrinsic Gaussian curvature of the disclination, valid for all allowed angular momenta for molecular orbitals once contributions from Chern flow and localized bound charges are removed via a process of taking an appropriate fractional part. We show that this numerically/experimentally obtained fractional charge vs. intrinsic Gaussian curvature data can be described using a novel relation, Eq.~\eqref{eq-masterTGR}, and used to recover a quantized topological response constant, the `gravitational coupling constant', characterizing the topological phase under investigation.}

\textcolor{black}{\emph{Wen-Zee gravitational response:} } Using Chern-Simons quantum field theories in  curved space, small real space deformations were previously shown to elicit a local \textcolor{black}{`gravitational' response} in continuum quantum Hall states on smooth curved manifolds \cite{1992-wen-fk}:
\begin{align}\label{eq-WZ}
\d\rr(\mbx)  = \rr(\mbx) - \rr_{0} = -e \frac{\k}{2\p}K_{G}(\mbx).
\end{align}
In the preceding equation,  $\d\rr(\mbx)$ is the excess charge density and $K_{G}(\mbx)$ is the local Gaussian curvature. The dimensionless linear response coefficient, $\k$, is the gravitational coupling constant (GCC).  This general prediction has since been shown to be valid for the specific case of continuum isotropic quadratically dispersing integer quantum Hall states on smooth manifolds \cite{1994-iengo-fk}. Integrating Eq.~\eqref{eq-WZ} over a closed manifold and using the Gauss-Bonnet theorem:
\begin{align}\label{eq-WZintegrated}
N_{e} - \n N_{\f} = 2\k (1-g),
\end{align}
where $g$ is the genus of the manifold, an integer; $N_{e}$ is the total number of electrons, an integer; $N_{\f}$ is the number of flux quanta piercing the surface, also an integer due to magnetic monopole charge quantization\cite{1931-dirac-em}; and $\n$ is the Hall conductance of the quantum Hall state in units of the conductance quantum, a  rational fraction. Substituting $g=1$ in Eq.~\eqref{eq-WZintegrated}, $2\k \equiv \n \mc{S}$ is the excess charge on a sphere, when compared with flat space with the same area. $\mc{S}$ is termed the `shift' of the quantum Hall state\cite{1992-wen-fk,1983-haldane-rt}. A remarkable insight follows: since Eq.~\eqref{eq-WZintegrated} is a linear relationship in $\k$ involving integers and a rational fraction, the GCC has to be a rational fraction. Thus, a significant physical consequence is that $\k$ cannot change continuously and is quantized in continuum quantum Hall states  \cite{1992-wen-fk}.

Using the fundamental physical principles of the discreteness of charge and gauge invariance, one can provide a succinct argument for the quantization of the Hall conductance, the celebrated eponymous characteristic of quantum Hall states. Crucially, these insights facilitated prediction of quantization of Hall conductance in \emph{all} gapped insulators \cite{1985-niu-nr,1982-thouless-fr,1988-haldane-eu}. In contrast, the phenomenology of gravitational response has yet to be generalized to a broader class of materials than just the continuum quantum Hall states. A valuable step in the direction was provided by our previous formulation of the problem for quantum Hall states on lattices\cite{2016-biswas-kx}. Substantially building on extant results, herein we broadly formulate the phenomenology of gravitational response for lattice topological quantum states. \textcolor{black}{We then} prove the quantization of the GCC and demonstrate its existence, i.e., show that it has nontrivial nonzero values in specific cases. We term this general quantized gravitational response as the ``topological gravitational response'' (TGR).

\textcolor{black}{\emph{Wen-Zee-like response on the lattice:} } In the context of lattice states of matter, disclinations provide a natural route to introducing spatial curvature and characterizing the gravitational response\cite{2016-biswas-kx}. The Gaussian curvature of the disclination is localized at the disclination core and thus singular. The \textcolor{black}{total instrinsic} curvature \textcolor{black}{of a disclination}, characterized by the extent to which Euclid's 5th postulate is violated as elucidated in Fig.~\ref{fig-disclinations}\textcolor{black}{, is equal to the Frank angle, quantized in the presence of local rotational symmetry}. Previously we showed that singular spatial curvature induced at a lattice disclination can be leveraged to characterize the gravitational response of quantum Hall states on the lattice\cite{2016-biswas-kx}. \textcolor{black}{This} approach \cite{2016-biswas-kx,2019-he-bc,2019-ozawa-jq,2016-hung-hc,2019-lu-kt} allows clear separation of the desired curvature-induced phenomena from spurious causes, such as arising from defects and disorder inadvertently introduced in constructs of slowly-varying small curvature on the lattice, which attempt a literal actualization of the continuum picture \cite{1992-wen-fk}. For quantum Hall states on crystalline lattices, we reported that each Landau level loses (an integer number of) states bound to such a disclination\cite{2016-biswas-kx}. 

We can succinctly summarize \textcolor{black}{gravitational response} for \textcolor{black}{lattice} quantum Hall states\cite{2016-biswas-kx} using a modified version of Eq.~\eqref{eq-WZ}, which captures gravitational response in both continuum and lattice quantum Hall states:
\begin{align}\label{eq-WZnew}
\text{frac}(-Q/e^{*}) = \text{frac}\le(\frac{\k}{2\p}\cdot\frac{e}{e^{*}}\cdot K_{T}\ri),
\end{align}
wherein $Q$ is the excess charge accumulated at the disclination; $\k$ is the GCC; $K_{T}$ is the \textcolor{black}{\emph{total} }integrated Gaussian curvature, simply equal to the Frank angle of the disclination; and $\text{frac}(x) = x - \lfloor x \rfloor$ is the fractional part of $x$. $e^{*}$ is the elementary unit by which charge can be locally modified in the material under consideration. For \textcolor{black}{continuum} quantum Hall states, $e^{*}$ is \textcolor{black}{simply equal to} the elementary quasiparticle charge (for instance, $e^{*}=e$ for integer quantum Hall states and $e^{*}=e/3$ for the $\nu=1/3$ Laughlin state).

Building on this formulation, here we provide the general results for topological gravitational response, in turn, first for all lattice insulators, and next, for Chern and related insulators. Consider three elementary charges associated with a specific gapped phase on the lattice: the electric charge, $-e$; the quasiparticle charge, $q_{p}e$; and the charge per lattice site, $q_{s}e$. $q_{p}$ and $q_{s}$ are expected to be rational fractions. The excess charge at a disclination is universal modulo integer multiples of $q_{p}e$ and $q_{s}e$, since these are charges associated with bound states and vacancies, respectively. Thus, if $e^{*} = q e$ is the greatest fractional charge such that $q_{p}e$ and $q_{s}e$ are integer multiples of $e^{*}$, then the excess charge at a disclination is determined modulo an integer multiple of $e^{*}$. Assuming local response, we associate an excess charge $Q_{v}$ with disclination type $v$. Now, considering all possible \emph{closed} manifolds formed by the lattice with well-separated disclinations, for each closed manifold the total electronic charge is:
\begin{align}
N_{e} = \sum_{v} n_{v} Q_{v} q + N_{p}q_{p} + N_{s}q_{s} + Q_{B},
\end{align}
wherein $n_{v}$ is the number of disclinations of type $v$, $N_{p,s}$ are integers, and $Q_{B}$ is the net charge from all flat regions in units of $e$. $Q_{B}$ is given by the number of sites times $q_{s}e$ for topological bands. (For the specific case of quantum Hall states, $Q_{B} =  \n N_{\f}$.) Thus, all terms other than the $Q_{v}$'s are known to be integers or rational fractions. Since there are infinite such equations, corresponding to realizations of all allowed manifolds, we have thus established that $Q_{v}$'s are also rational fractions. If $Q_{v}$ (modulo $e^{*}$) is linearly related to the integrated curvature of the disclination in units of $2\pi$, i.e., a rational fraction, then the slope $\k$, i.e., the GCC, must thus also be a rational fraction. Thus we arrive at the remarkable conclusion that the gravitational coupling constant (GCC), if non-zero, must be quantized, and is a characteristic property of the gapped topological phase.

\begin{figure}[t]
\begin{center}
\resizebox{0.8\textwidth}{!}{\includegraphics{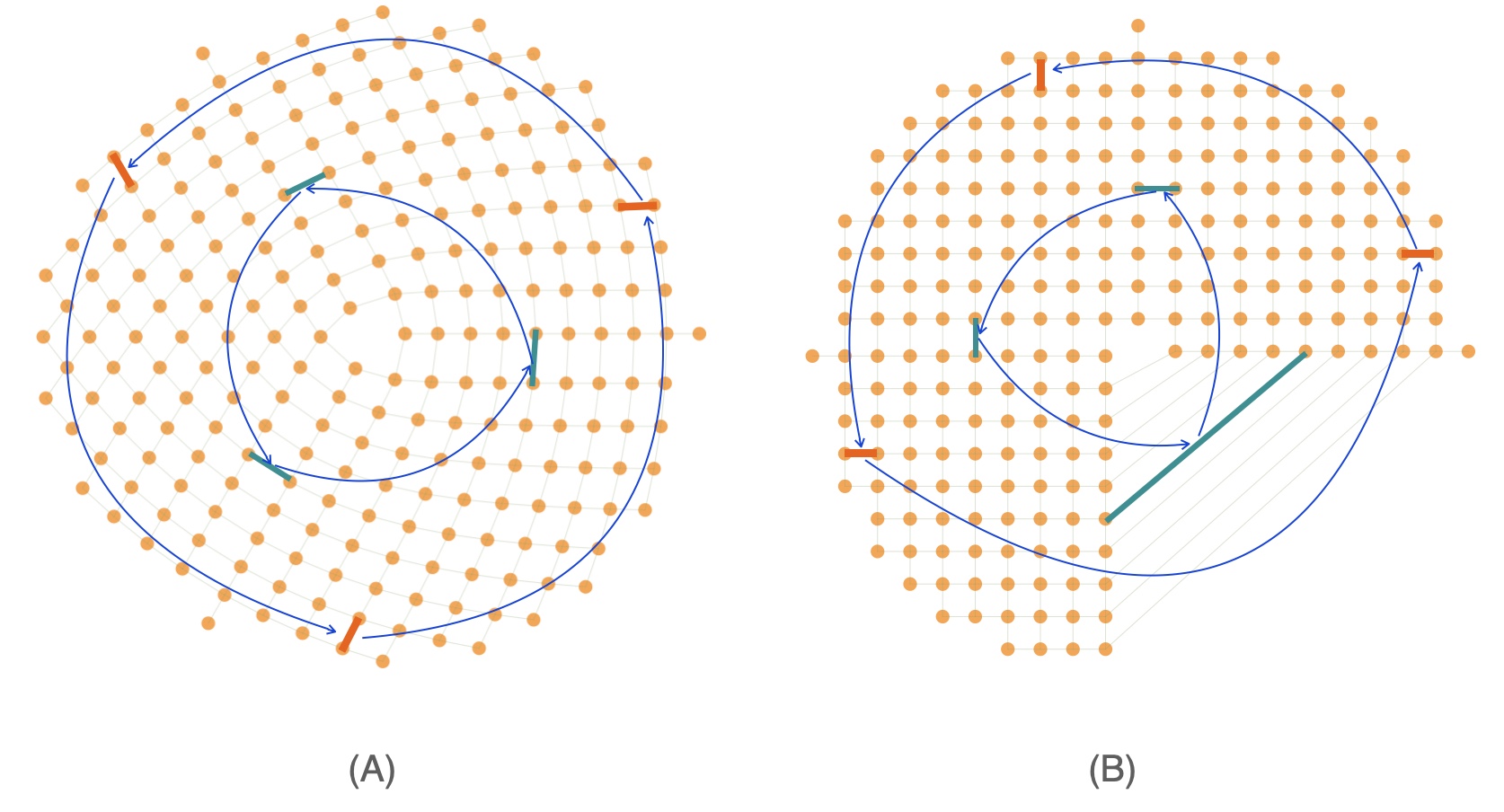}}
\caption{\emph{General prescription for constructing Hamiltonians at disclinations.} A three-fold disclination on a square lattice, (A) shown using conventional view and (B) shown as arising from the flat space lattice by replacing the last plaquette with bonds. Using rotation invariance of the Hamiltonian in flat space, we can deduce unitary transformation $U$ corresponding to elementary rotation up to an overall phase, $\y$. Using this unitary transformation (blue arrows), flat space Hamiltonian on first plaquette can be used to construct Hamiltonian on any other plaquette in the presence of disclination (bonds related by rotation are indicated). Final set of bonds sewing the final plaquette to the first (long bonds in (B)) are constructed from bonds connecting second and first plaquettes. Details in \cite{tgr-sm}.}
\label{fig-sewing}
\end{center}
\end{figure}

\textcolor{black}{\emph{Topological Gravitational Response:}} We now formulate TGR in two dimensional band insulators with $n$-fold rotation symmetry, $R_{n}$, by generalizing the construction of Hamiltonians at disclinations \cite{1997-tamura-yf,2013-ruegg-ep,2018-he-ff,2019-liu-fm,2020-li-gl} (see Fig.~\ref{fig-sewing}, details in \cite{tgr-sm}). The unitary operation, $U$, which denotes the action of rotation $R_{n}$ on the internal Hilbert space at each lattice site, is obtained up to an overall phase $e^{i\y}$  by requiring that $R_{n}$ commute with the family of Hamiltonians representing the insulator phase. Since $\le(R_{n}\ri)^{n}$ is fixed by the boson/fermion/anyon nature of the Hilbert space\footnote{For e.g., in a system with two orbitals per site, the free-space rotation representation acting on the orbitals may be a direct sum of two one-dimensional representations, or an irreducible two-dimensional spin-$1/2$ representation. In the first case, rotation by $2\p$ gives $1$, which we call the `bosonic' character of the Hilbert space. In the second scenario, rotation by $2\p$ gives $-1$, which we call `fermionic'. Looking ahead, in more exotic correlated systems with anyonic quasiparticles, rotation by $2\p$ may give some intermediate phase, which we call `anyonic'.}, $\y$ must be equal to $2\p k/n$ with $k=0, 1 \ldots (n-1)$. For `layered' phases\textcolor{black}{, e.g., composed of independent Chern insulator bands}, multiple $\y$'s exist. Since an $m$-fold disclination is composed of $m$ elementary lattice wedges and $e^{i\y}$ can be viewed as a Peierls phase acquired every time an electron hops from one wedge to the other,  there is a fictitious out-of-plane magnetic flux of $- m \y/(2\p)$ quanta (with an unknowable offset), associated with the disclination. Thus, adiabatically changing $\y$ from an allowed value $\y_{1}$ to another $\y_{2}$ will lead to an \textcolor{black}{inflow} of charge $-Ce$ per flux quantum at the disclination, i.e., $m C e (\y_{2}-\y_{1})/(2\p)$ in total. \textcolor{black}{Note that during this adiabatic process rotational symmetry is broken because of disallowed values of $\y$, however the gapped topological phase remains intact since, as assumed, its integrity does not depend on the presence of rotational symmetry.}

Incorporating this physics and setting $\y = 2\p k/n$, the phenomenon of topological gravitational response (TGR) in lattice topological systems with $n$-fold rotational symmetry and Chern number $C$ is given by the following expression for the fractional charge accumulated at an $m$-fold disclination:
\begin{align}\label{eq-masterTGR}
\text{frac}(-Q_{m}/e^{*}) &= \text{frac}\le[\le(\frac{e}{e^{*}}\ri)\le(\k\le(1-\frac{m}{n}\ri) - \frac{m C k}{n}\ri)\ri]\nn\\
&= \text{frac}\le(\tilde{\k} - \le(\tilde{\k} + \tilde{C} k\ri)\frac{m}{n}\ri),
\end{align}
wherein $\tilde{\k} = \le(e/e^{*}\ri)\k$ and $\tilde{C} = \le(e/e^{*}\ri)C$. When $C\neq 0$, one must take into account the fact that a given topological phase can be labeled by \emph{any} integer value of $k$ (noting that the choice of $\y=0$ is arbitrary and depends on the gauge for $U$). For simplicity, in what follows we will assume that $e/e^{*}$ (and therefore, $\tilde{C}$) is an integer, which holds for typical scenarios. 

\begin{figure}[h]
\begin{center}
\resizebox{0.8\textwidth}{!}{\includegraphics{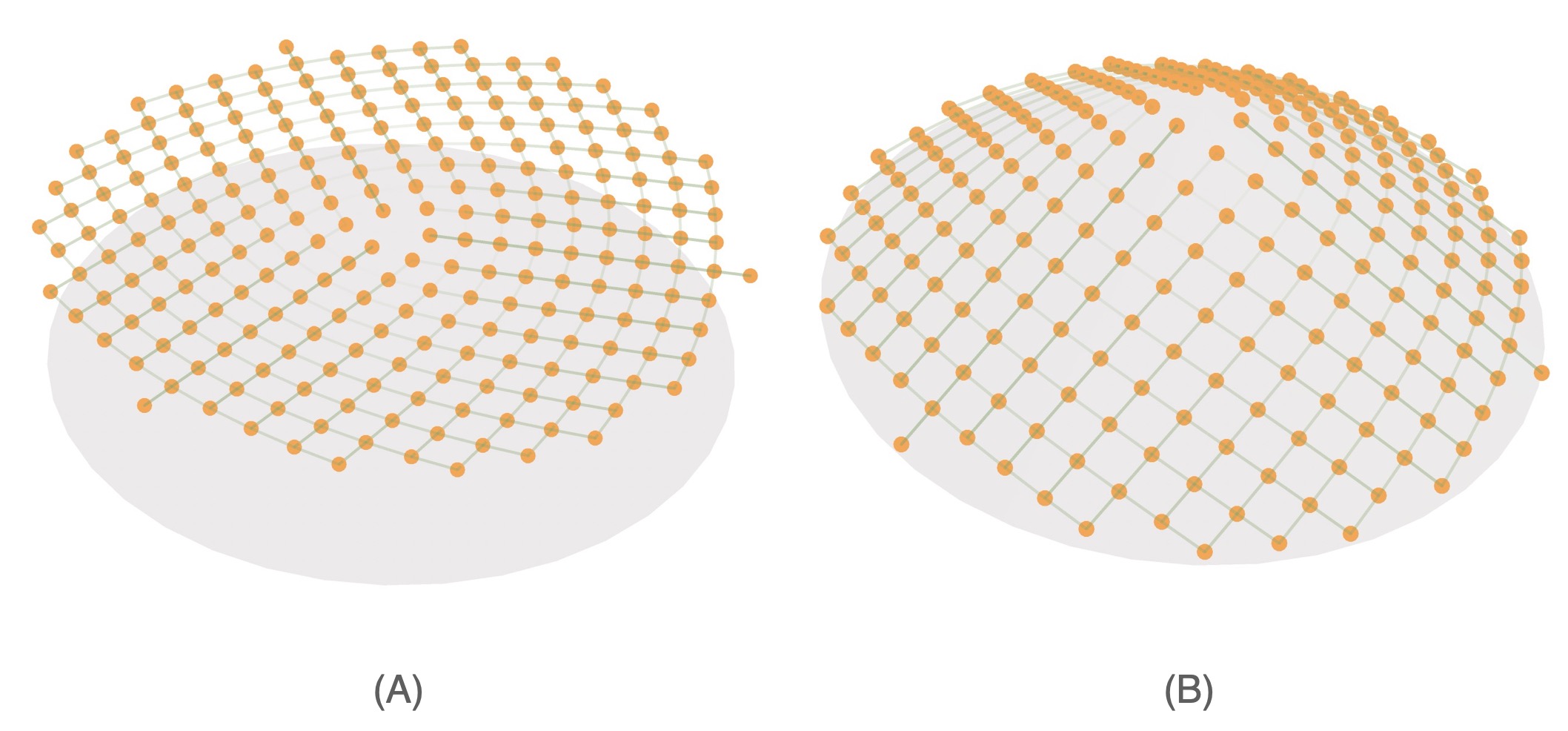}}
\caption{\textcolor{black}{\emph{Disclinations with same intrinsic (Gaussian) and differing extrinsic curvatures.} A three-fold disclination on a square lattice, with the same intrinsic geometry as the ones shown in Figure~\ref{fig-disclinations}, can be affected in distinct ways by strain. For example, laying it flat, (A), or letting it relax into a conical geometry, (B), results in different strain profiles and corresponding variation in the strengths of electronic hopping amplitudes (represented above by varying bond opacities). Electronic motion is thus affected by this `extrinsic' geometry\cite{2022-vafa-sf}. However, when the insulating energy gap is preserved, the total accumulated fractional charge depends only on the Gaussian curvature and is the same for both strain cases shown here, equal to that obtained for the strain-agnostic case sketched in Figure~\ref{fig-disclinations}(A). For details, see \cite{tgr-sm}.}}
\label{fig-strained}
\end{center}
\end{figure}

Owing to the relating of the fractional parts of the quantities on both sides of Eq.~\eqref{eq-masterTGR}, the value of $\tilde{\k}$ can only be reported modulo the greatest common divisor of $n$ and $\tilde{C}$ (in which we define a divisor to be a number arrived at via division by a natural number). $\k$ is deduced from $\tilde{\k}$ by dividing by $e/e^{*}$. The remarkable result thus obtained is that the value of $\tilde{\k}$ calculated as prescribed is \emph{universal}. In other words, given a topological model, the quasiparticle charge and behavior under $2\p$ rotation (i.e., whether bosonic/fermionic/anyonic), $\tilde{\k}$ is the same for all classes of disclinations, independent of the details of how the disclination core is patched up; independent of local disorder; \textcolor{black}{and also independent of long range strain arising from the specific extrinsic curvature effects related to how the disclination is embedded in space (see Figure~\ref{fig-strained}, for details see \cite{tgr-sm}).} 

\begin{center}
\begin{table}
\begin{center}
\begin{tabular}{|l||l|c|c|c| c|c|c|} 
\hline
Model &  $n$ &$ e^{*}/e$ &  $C$ &$ \tilde{C} $& $\tilde{\k} $& $\k$ \\ 
\hline
\hline
QWZ & $4$   & $1$ & $ \pm 1$ &$ \pm 1 $& $ 1/2 \mod 1 $& $1/2 \mod 1$ \\ 
\hline
QWZ 2 & $4$ &$ 1 $ & $ 2 $ &$ 2 $& $ 0 \mod 2 $& $0 \mod 2$ \\ 
\hline
Sticlet & $6$ &$ 1 $ & $ \pm 2$ &$ \pm 2 $& $ 0 \mod 2 $& $0 \mod 2$ \\ 
\hline
Alase-Feder & $6$ &$ 1 $ & $ \pm 3$ &$ \pm 3 $& $ 3/2 \mod 3 $& $ 3/2 \mod 3$ \\ 
\hline
Haldane &  $6$ &$ 1/2 $ & $ \pm 1$ &$ \pm 2 $& $ 1 \mod 2 $& $1/2\mod 1$ \\ 
\hline
Kane-Mele & $6$ &$ 1$ &  $\pm2^{*}$ & $\pm2^{*}$ & $ 0 \mod 2$& $ 0 \mod 2$ \\ 
\hline
\end{tabular}
\caption{\emph{Tabulation of numerically calculated GCCs for specific topological band insulators, establishing universal phenomenology of TGR.} All quantities are defined in the context of Eq.~\eqref{eq-masterTGR} in the main text. See accompanying text for import of these values. The models, \textcolor{black}{defined in detail in \cite{tgr-sm},} are adopted from the following references: QWZ/QWZ2\cite{2006-qi-xm}, Sticlet\cite{2012-sticlet-kq,2019-leonforte-dg}, Alase-Feder\cite{2021-alase-fp}, Haldane\cite{1988-haldane-eu}, Kane-Mele\cite{2005-kane-qq,2005-kane-xe}. $^{*}$For the Kane-Mele model, a topological insulator constructed from two elementary time-reversed copies of Chern insulators, we use the sum of absolute values of their Chern numbers.}
\label{resultsTable}
\end{center}
\end{table}
\end{center}

To demonstrate the existence of such nontrivial topological gravitational response, we have \textcolor{black}{numerically} evaluated the \textcolor{black}{gravitational coupling constants} (GCCs )for some well-known examples of topological band insulators such as Chern insulators and time-reversal-invariant topological insulators. The results are summarized in Table~\ref{resultsTable}. We checked fractional charges at disclinations with between $1$ and $\sim 10$ wedges, centered at different symmetry points belonging to different classes \cite{2013-gopalakrishnan-lk} and with different allowed values of the overall phase factor, $e^{i\y}$. For each model, our numerical results for all such disclinations agree with the general theory presented with  Eq.~\eqref{eq-masterTGR}, yielding the GCC values in Table~\ref{resultsTable}. Details are provided in \cite{tgr-sm}.

\begin{figure}[t]
\begin{center}
\resizebox{0.8\textwidth}{!}{\includegraphics{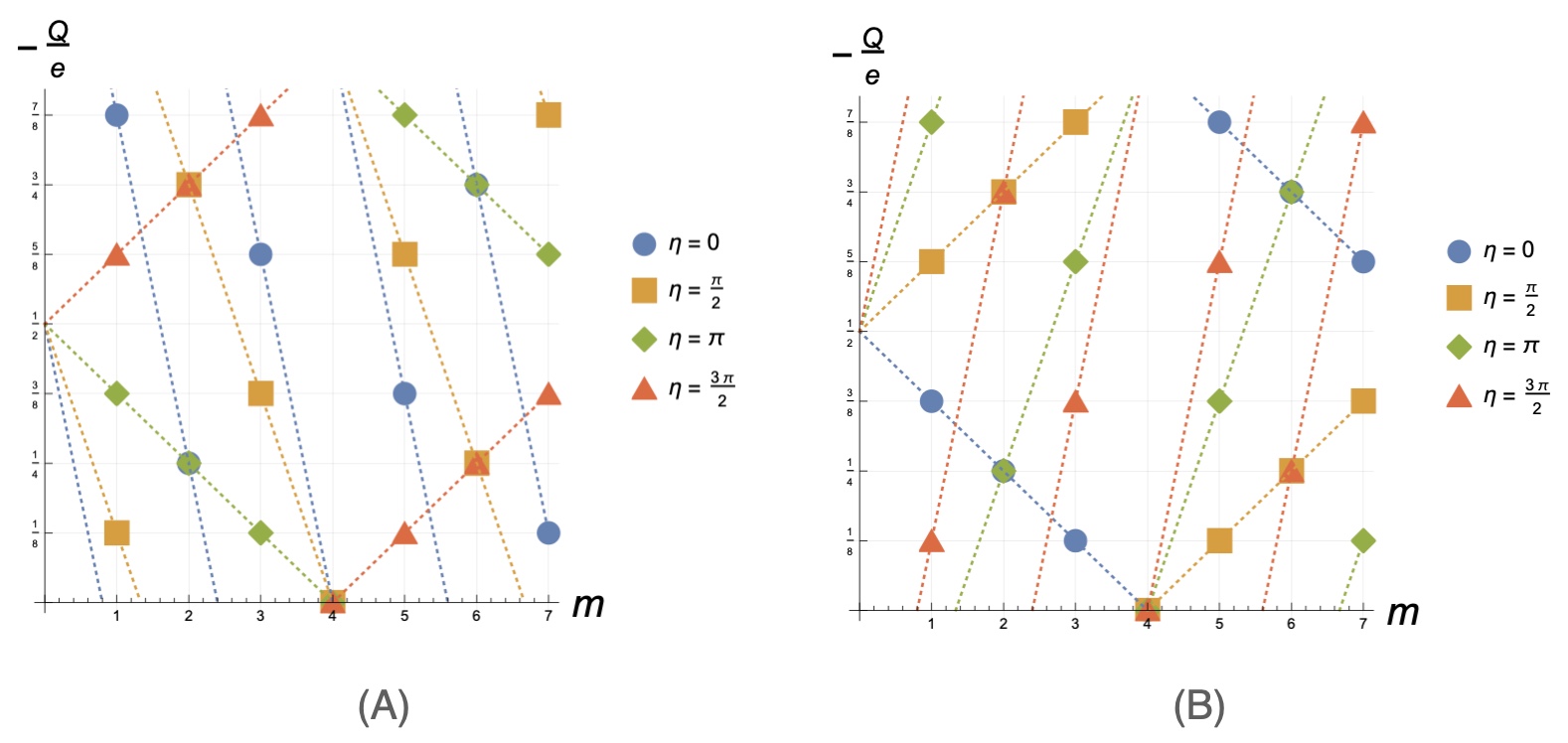}}
\caption{\emph{Calculating the GCC.} Fractional charges calculated in the $C=-1$ phase of the QWZ model at $m$-fold disclinations on a square lattice, centered at (A) the plaquette center and (B) at a lattice site. Different keys refer to allowed values of the phase ambiguity $\y$. Dashed lines are fits using Eq.~\eqref{eq-masterTGR}, yielding $\k = 1/2, 5/2$ for (A,B) respectively. The GCC can only be reported modulo the greatest common factor of $n=4$ and $C=-1$, which is $1$. Thus, the \emph{universal} value of the GCC is $\k=1/2 \mod(1)$ for this topological phase.}
\label{fig-GCCcalculation}
\end{center}
\end{figure}

Briefly, our procedure for extracting the GCC from numerical data is exemplified in the following context: consider the $C=-1$ phase of the QWZ model\cite{2006-qi-xm} on the $n=4$ fold symmetric square lattice. $U = e^{i\y}\text{diag}(1,-i)$, acting on the two dimensional internal Hilbert space of each lattice site, represents the unitary transformation corresponding to clockwise rotation by $\p/4$. The overall phase, $\y$, is a multiple of $\p/4$, assuming $U^{4}=1$. $e^{*}/e$ is equal to $1$. On the square lattice, there are two nonequivalent centers of $4$-fold rotational symmetry: the lattice site and the plaquette center (Fig.~\ref{fig-disclinations}). Fig.~\ref{fig-GCCcalculation} shows our numerical evaluation of the fractional charge accumulated at these two families of disclinations, for different values of $\y$. The \emph{universal} value of the GCC is $\k=1/2 \mod(1)$. (This result is also consistent with TGR at disclinations centered on the bond centers, not shown for brevity.)

\textcolor{black}{\emph{Discussion:}} An unexpected relation is revealed by our results in Table~\ref{resultsTable}: the GCC is $1/2$ (equivalently, $-1/2$) times the Chern number, modulo the greatest common divisor of the Chern number and $n$ (the lattice has $n$-fold rotation symmetry). Our observation is consistent with many long-established results in the continuum ($n=\infty$). Continuum Landau levels, indexed as $k=1, 2, \ldots$, exhibit GCC values $\k = k - 1/2$\cite{1992-wen-fk}. The $k$-dependence can be explained using the cyclotron orbit angular momentum $L_{k} = k \hbar$ \cite{2002-yoshioka-zl,2007-jain-fk}, which implies $\y n = -2\p k$. Since each Landau level has a Hall conductance corresponding to $C=1$,  the GCC for the family of Landau levels is equivalent to $\k = 1/2 \mod(1)$ in agreement with the known pattern above. Remarkably, model Laughlin states are also characterized by $\k=1/2$ \cite{1992-wen-fk}. This value is consistent with the GCC vs.\ Chern number pattern noted previously if we interpret the Laughlin state as a single filled Landau level of composite fermions \cite{2007-jain-fk} (with $e^{*} = \nu e$). We note that the GCC is doubled (and becomes equivalent to $0$) for time reversal invariant topological insulators formed from pairing time-reversed Chern insulators, for which the right hand of Eq.~\eqref{eq-masterTGR} becomes effectively proportional to $m$, i.e., one simply observes that there is a fixed charge `per wedge' \cite{2019-liu-fm,2020-li-gl}.  

\textcolor{black}{We have used disclinations that, due to the presence of rotational symmetry, allow controlled creation of pristine spatial curvature free of undesirable  interference effects on electronic motion, to distill the universal phenomenon of topological gravitational response, viz., the charge response to \emph{intrinsic} spatial Gaussian curvature. This physics is now ready for incorporation in other scenarios where gravitational response acts in conjunction with other effects. A notable example is the physics of anomalous viscosity in continuum Hall states, where the anomalous viscosity\cite{1995-avron-fk,2007-tokatly-fp,2009-tokatly-fk}, proportional to the Wen-Zee gravitational constant in the presence of continuous rotation symmetry\cite{2009-read-fk}, shows up in current and charge responses to nonuniform electric fields in conjunction with other causes such as the Hall effect and effects arising from the electrodynamics of continuous media\cite{2012-hoyos-fk,2010-taylor-kq,2021-chen-xl}. What happens to this physics in lattice systems\cite{2021-kozii-hs} as considered herein? These questions may serve as promising starting points for future inquiry.} 

We conclude by drawing attention to an intriguing point: our results demonstrate linear relationship between charge and net curvature, without observable deviations from linearity. What are the implications of absence of nonlinear terms in the response? Exploring the consequences is a promising future avenue of research. Another exciting line of inquiry is to establish if aspects of the universal response to gravitational perturbations reported here can be further generalized to apply to analogous soft matter systems such as in \cite{2022-zhang-mr}. Finally, there is the exciting possibility of explicating general connections between topological quantum field theories in the presence of gravity and the well-defined analytically tractable quantum lattice models studied in this work.

\emph{Acknowledgements}: We thank David Nelson, Bert Halperin, Michael Stone, Tony Zee, Steve Kivelson, Cumrun Vafa, Xiao-Gang Wen, and Yichen Xu  for useful discussions. We thank the Purdue Research Foundation, Purdue University Startup Funds and the Bilsland Fellowship for financial support. This research constitutes a substantial portion of the dissertation work reported in \cite{2021-jiang-cy}.

\emph{Author contributions}: RRB conceived of, designed and actualized the research; GJ and RRB performed calculations; GJ, YK, SI-B, and RRB checked calculations and discussed the results; GJ, SI-B, and RRB wrote the paper.

\end{document}